# Data-centric challenges with the application and adoption of artificial intelligence for drug discovery

Ghita Ghislat[a], Saiveth Hernandez-Hernandez[b], Chayanit Piyawajanusorn[c], Pedro J. Ballester[c*]

[a] Department of Life Sciences, Imperial College London, London SW7 2AZ, UK.

[b] INSERM U1068, Marseille 13009, France.

[c] Department of Bioengineering, Imperial College London, London SW7 2AZ, UK.

**\* Corresponding author:** Pedro J. Ballester (p.ballester@imperial.ac.uk)

**Abstract**

**Introduction:** Artificial intelligence (AI) is exhibiting tremendous potential to reduce the massive costs and long timescales of drug discovery. There are however important challenges currently limiting the impact and scope of AI models.

**Areas covered:** In this perspective, the authors discuss a range of data issues (bias, inconsistency, skewness, irrelevance, small size, high dimensionality), how they challenge AI models, and which issue-specific mitigations have been effective. Next, they point out the challenges faced by uncertainty quantification techniques aimed at enhancing and trusting the predictions from these AI models. They also discuss how conceptual errors, unrealistic benchmarks and performance misestimation can confound the evaluation of models and thus their development. Lastly, the authors explain how human bias, whether from AI experts or drug discovery experts, constitutes another challenge that can be alleviated by gaining more prospective experience.

**Expert opinion:** AI models are often developed to excel on retrospective benchmarks unlikely to anticipate their prospective performance. As a result, only a few of these models are ever reported to have prospective value (e.g. by discovering potent and innovative drug leads for a therapeutic target). The authors have discussed what can go wrong in practice with AI for drug discovery. We hope that this will help inform the decisions of editors, funders investors and researchers working in this area.

## 1. Introduction

A 2023 study[1] has estimated that the cost of developing a new drug has risen to a staggering average of US$6.16 billion. This represents a massive increase over previous estimations, e.g. this 2016 study[2] where the average cost of a new drug was found to be US$2.87 billion, which in turn was a higher cost than previous studies after correcting for inflation.

While clinical studies represent the most expensive part of drug development, most time-saving and cost-saving opportunities are in the earlier discovery and preclinical stages[3]. In addition, preclinical efforts themselves account for more than 43% of expenses in pharma[4], this enormous aggregated cost driven by the high percentage of projects that fail at every stage, including hit identification and lead optimisation. Pharmaceutical companies rarely disclose the latter attrition rate across projects. This GSK study[5] is an exception and reported that 93% of their antibacterial projects did not achieve any drug lead despite massive investments.

A range of reasons have been proposed for why research and development (R&D) productivity in drug discovery is so low[6]. Big pharmaceutical companies have been found to be generally much less productive than small but more agile players such as biotechnology companies[1]. Lower productivity has been pointed out as the reason why most drugs achieving regulatory approval are currently not discovered by big pharma[7], but bought by them to push drugs to the market using their vast resources and unique clinical trials know-how.

Given the substantially higher R&D productivity outside big pharma, increasingly more partnership deals are made with biotech companies and academic drug discovery centers[8,9]. Artificial intelligence (AI) startups and biotechs have inked many of these profitable deals based on their early research outcomes[10]. These successes suggest a tremendous potential of AI to increase R&D productivity in small-molecule drug discovery, which is starting to be evident in the public domain[11,12,21,13–20]. Ongoing research on how to best apply AI algorithms to these problems guided by multidisciplinary teams will eventually lead to the full realisation of such potential[22]. By contrast, no matter how strong the computer science expertise leading them is, AI for Science projects focusing on the development of new algorithms will be sterile if it is guided by benchmarks that are not well aligned with their drug discovery projects[23].

Despite these highly encouraging facts, there are important roadblocks ahead, little known by many evaluating, investing or even working on AI for drug discovery, which greatly limit its potential. Often these roadblocks will be more specifically presented in the context of Machine learning (ML), the most developed branch of AI. Most of these challenges are common to other ML applications[24,25]. However, they tend to manifest themselves in different ways for different drug discovery problems and often need problem-specific mitigations. Figure 1 summarises the challenges we will overview in this paper, which come from data issues, uncertainty quantification, model evaluation and researcher bias.

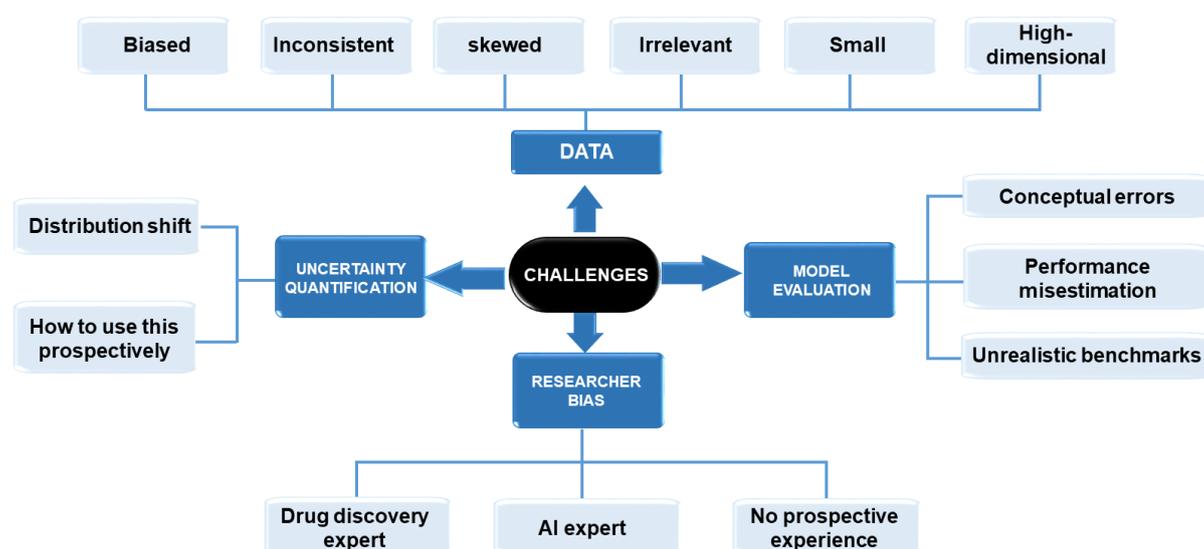

**Figure 1.** Overview of AI challenges, some of them specific to particular drug discovery problems.

## 2. Challenges from data issues

## 2.1 Biased data

Let us consider a perfect dataset whose instances all have labels that were measured without errors using exactly the same protocol. Even in this idealised situation, such instances could unevenly sample the label distribution and hence any model built on this dataset will be biased towards this region of the distribution. This means that the model could generalise poorly to the unseen regions of the distribution. As an example, think of models trained to predict the activity of small-molecule drugs on their cognate targets. These models are more likely to be predictive on other drugs than on molecules that are still to be tested on any target. Sample bias can also occur with the joint distribution from which feature vectors are sampled and furthermore such bias is convoluted with label bias[26,27]. In practice, training set and test set will generally be biased in different ways. This is called the distribution shift: the larger this shift is, the more unlikely will the model be to generalise[28,29]. The impact of a shift can be studied by applying a suitable training-test splitting strategy[30,31].

## 2.2 Inconsistent data

Data instances can also be inconsistent if their labels are not measured in the same way. Models are typically trained with data generated by different laboratories. Sometimes even small differences in the data-generating protocol, such as variations in equipment calibration or sample preparation methods, can result in poor generalisation of the models trained on such heterogeneous datasets. For instance, when the activities of the same drug molecules on the same cancer cell lines are measured *in vitro* in different laboratories using the same protocol, strong differences have been observed in many cases[32]. These inconsistencies, driven by a variety of factors[33,34], can be reduced by standardising practices across laboratories[34–36].

Inconsistencies can also be detected by testing a model trained on data from one laboratory using data from another laboratory. When the data is inconsistent, the performances of such models will be much worse than those from models trained and tested on data from the same laboratory[37]. This type of label measurement error will generally be convoluted with data biases and other challenges posed by data. Common practice in relevant communities such as computational chemistry or chemoinformatics is restricting to the highest quality data to reduce inconsistencies, but this usually reduces the accuracy and applicability of AI models. Alternatively, one can gather a larger training set by merging activity-labelled molecules from closely related targets[38], adding molecules whose target activity is measured with surrogates such as target binding[39,40], employing features from lower-quality but diversity-expanding protein-ligand complexes[40] or generating large volumes of assumed target inactives[41]. These larger datasets hence come at the cost of increasing inconsistencies between data instances, which is justified if it leads to better generalisation and/or generality of the models trained on them (this can be hard to anticipate from analysing inconsistencies). Inconsistencies can also be artifacts arising from problem-misaligned similarity metrics[42].

## 2.3 Skewed data

A particularly challenging subtype of biased datasets are those with skewed distributions of label measurements. When the label is binary, such datasets are called class imbalanced[43]. Extreme class-imbalanced datasets are the norm in early drug discovery, where molecules with affinity for a target (the minority class) are often modelled to be 50 times less frequent than likely or confirmed inactives (the majority class)[44]. As extreme as this simulated 2% hit rate is, screened libraries are even more imbalanced (their hit rates are typically between 0.01% and 1%[45]). Models trained with many decoys (i.e. likely-inactive molecules often designed to be hard to distinguish from the known actives for the considered target) have been shown to reduce false positives across molecular targets[41,46–48]. Among other approaches that can be employed to mitigate class imbalance in classification, you have active learning[49], which has long been used for this purpose[50], oversampling[51] and semi-supervised learning[52].

**2.4 Irrelevant data**

One often has to select a set of features able to capture the causative factors of the property to predict (the label). Feature selection techniques[53] can retain at least some of the predictive features and discard the rest as likely irrelevant to the prediction[54]. Some retained features will be predictive of the label because they correlate with causative features, not because they are causative themselves. This is more likely to occur when there are few data instances and there are many eye-opening examples available (http://www.tylervigen.com/spurious-correlations).

A more concerning possibility is that some of the causative features are not modelled. This could happen because we do not know about them. But it also happens when we do, but there is not an effective way to calculate them. For instance, we know that the energetic contributions of ligand desolvation, pocket desolvation and induced-fit rearrangements prior to forming the protein-ligand complex are challenging to quantify accurately in a fast manner, which limits the accuracy of scoring functions[55,56]. In this particular problem, these factors are accounted for, to some extent, when training AI models across many complexes. This is because the algorithm learns how affinity differences relate to the same protein bound to different ligands and the same ligand complexed with different binding pockets, which comprise various protein conformations and desolvation differences.

**2.5 Small data**

Even if the data was unbiased, consistent, balanced and relevant, the number of samples could be too small for the supervised learning algorithm to infer all the information required to accurately predict other samples. This happens despite the tireless work of those building and maintaining relevant data resources[57,58]. Self-supervised learning could be the best approach in these cases. This essentially consists in pre-training a deep learning model using a large number of unlabeled instances and then tuning the pre-trained model with the much fewer labeled instances. For example, the deep learning model can be a large-language model pre-trained with 77 million SMILES sequences and tuned to predict a range of molecular properties such as molecule-induced toxicity[59,60]. The value of this type of approaches has already been shown prospectively for structure-based prediction of protein-ligand binding affinities[19].

Physics-informed methods such as docking have long been used to tackle a specific small-data problem: structure-based virtual screening. Indeed, docking can be applied to the discovery of small-molecule binders for a therapeutic target with few or even none of these binders known (such targets are often termed "undruggable") as long as the binding pocket is known or correctly predicted[61]. This transferring of what the method has learnt (docking is calibrated on activity/binding and structural data from various ligand-bound targets) to other ligand and target molecules is in ML called transfer learning[62]. ML has been extensively applied to enhance docking, e.g. to improve virtual screening hit rates across targets[63,64]. Rigorous retrospective validations show that some of these ML scoring functions are highly predictive across unseen test targets including those dissimilar to any training target[19,47,65–67]. Such level of generalisation to dissimilar protein-ligand complexes proves that ML scoring functions are able to learn the physics of protein-ligand binding from data better than other virtual screening method types (e.g. docking) even if this is not explainable to a human. Therefore, transfer learning models are also promising for targets without any training data[15,68].

Another approach that is gaining traction is to start by generating training data for the intended target. These datasets are often a medium-throughput *in vitro* screen of molecules for target activity. Successful studies in this category have used learning algorithms such as random forest[69] or directed message passing neural network[17,70] to train AI models on the generated datasets. The initial investment in the *in vitro* screen is fully justified by the subsequent fast and effective discovery of potential drug leads. The question remains, however, of whether iterative *in silico-in vitro* cycles would be even more effective.

## 2.6 High-dimensional data

This challenge is common in biomarker discovery[25,71]. This is the case of cancer drug response markers, where one aims at predicting the response of cancer patients to a drug treatment from the molecular profiles of their tumours[72–77]. For example, one could have only 100 tumours of a given cancer type, but each tumour be described by over 400,000 features (the DNA methylation profile). A key family of techniques to face this challenge is feature selection[78]. Accurately estimating how well the model will perform in unseen datasets is further challenged by considering many features and requires careful bootstrapping or nested cross-validation[77].

## 3 Challenges from uncertainty quantification

Quantifying the uncertainty of a prediction, i.e. how reliable it is, introduces another dimension to decision making. It is no longer only about selecting the molecules with the most potent predicted activities to be tested *in vitro* on the target, but also to discard molecules whose predictions are unlikely to be correct. For example, Gaussian Process (GP)-based uncertainty prediction was employed to select a few likely and potent small-molecule inhibitors of several molecular and phenotypic targets[79]. Following *in vitro* tests, some of these molecules were found to exhibit nanomolar affinity for these targets [79].

Conformal Prediction (CP) is an attractive alternative to GP in that CP is a model-agnostic uncertainty quantification framework and thus can be combined with any learning algorithm[26]. For instance, CP-enhanced regression models have already provided accurate error bars for predicted $pGI_{50}$s of molecules across NCI-60 cancer cell lines[80]. This is also the case with strong distribution shifts[28], although it was not guaranteed, as these shifts break the CP-required assumption of exchangeability (data distributions are different beyond allowed permutations)[81]. While individual prospective validations of CP models have not been reported yet, some studies have argued a prospective benefit of CP in combination with other virtual screening methods[82,83]. Future prospective studies can define the most reliable predictions as those with the smallest CP error bars, among other meaningful definitions. Note that restricting to the most reliable predictions can discard the molecules predicted to be most potent. In this context, one could also wish to test *in vitro* the predictions with the largest CP error bars (i.e. the riskier ones), since retraining the model including these results could be the most effective way to improve its performance on such high-uncertainty chemical space regions. The use of CP-based classifiers is not as intuitive as in regression, thus requiring more reflection[84].

## 4 Challenges from model evaluation

### 4.1 Conceptual errors

While there are many conceptual errors that could be addressed, we will focus on the popular misconceptions around the impact of overfitting on model generalisation. Let us consider a problem where a random forest model is predictive on a test set whose instances were not used in any way to train or select the model. The same model is even more predictive on the training set. We have witnessed very senior domain experts claiming that such model would not be predictive, even though the results conclusively show that this is the case (note the absence of information leaking of any kind). A valid form to express this conceptual error would instead be: "if your model overfits the training data then it will generalise poorly to other datasets".

To see why this claim is incorrect in general, let us start with the operational definition of overfitting, as the situation where the model performs much better on the training set than on the test set. This reflects the fact that overfitting is not a binary property but a continuous one, as it occurs in different degrees. The first issue with this claim is hence that model overfitting is ill-defined, as the same model will generally have different performances on different test sets and thus different levels of overfitting. Therefore, overfitting is not a property of the model, but a property of both the learning algorithm and the employed datasets. Overfitting cannot be avoided or removed, but it can be mitigated (e.g. by tuning regularisation hyperparameters via cross-validation of the training set and then retraining with the entire training set using the optimal values of these hyperparameters). More importantly, the second issue with this claim is that not only it has not been proven, but there exist numerical and analytical counterexamples[85].

### 4.2 Performance misestimation

Model performance is estimated using a set of benchmarks[61,86]. A benchmark is here a retrospective analysis of existing data that should aim at anticipating how well a model would perform prospectively on the considered therapeutic target. In the case of an ML model, there is also a need to define a training set for the same target, which can be carried out by partitioning benchmark data into a training and test set (e.g. by AVE split[86], scaffold split[23] or dissimilar-molecules split[87]). Multiple repetitions, each initialised with a different random seed, of a given split are required to estimate the variance in performance of a model. If comparisons are made based on a single split, this will ignite the suspicion of that split being the more favorable for the authors' narrative, especially when the difference between the compared models is small.

Other sources of misestimation can be employing convenient metrics, which do not align with the considered problem. This is easy to happen given the large number of available metrics independently introduced by the AI-precursor communities (statistical learning, data mining, information retrieval, pattern recognition, etc.) to be adapted to the characteristics of their problems. Some of these metrics were even rediscovered: MCC is known as the φ coefficient in Statistics and was introduced by Yule in 1912, but later rediscovered by Matthews in 1975 as the MCC[88]. In this context, the challenge is to find the most suitable metrics for each problem type. For instance, ROC-AUC has been promoted as a primary metric for virtual screening[89]. However, while this is a perfectly valid metric for many classification problems, ROC-AUC is not suited for class-imbalanced early-recognition problems such as virtual screening[90,91]. As intuitively shown in [90], if half of the actives are retrieved at the very beginning of the rank-ordered list and the other half at the very end, that method will obtain a ROC-AUC of 0.5. This is the same value that would be obtained by a method that randomly distributes all the actives across the ranks. Since more true actives would be found by in vitro testing the top molecules from the first method than by testing the top ones from the second, the first method would have better virtual screening performance. This and other misalignments led the authors to conclude that ROC-AUC is "clearly a bad metric to discriminate among VS methods" [90]. Metrics more aligned to virtual screening include the hit rate and the normalised enrichment factor[31].

### 4.3 Unrealistic benchmarks

Benchmark results can be used to select a method[47] and/or calibrate its settings[61] for prospective applications. The focus has traditionally been on building benchmarks that are hard for a particular method type, without evaluating how well this can anticipate prospective performance. For example, the original DUD[61] was built to prevent docking from artificially exploiting physicochemical properties that are only found in the minor class (actives) and not in major class (inactives). But this benchmark suite does not assess, for instance, how well the model generalises to dissimilar molecules[31]. Therefore, a model excelling at one of the DUD benchmarks is likely to struggle when screening an ultra-large compound library.

LIT-PCBA[86] provides a more realistic distribution of chemical properties as it is derived from high-throughput *in vitro* tests. A shortcoming is that, in generating LIT-PCBA, many molecules considered problematic were discarded and thus its chemical diversity is unrealistically low[86], in sharp contrast with the libraries that are screened prospectively[92]. Molecules were discarded using medicinal chemistry filters to remove likely frequent binders, but such filters have been found to wrongly discard even FDA-approved drugs[93,94]. Given their scarcity and high value, we recommend that active molecules are only discarded if there is certainty that there are no true binders. Inactive molecules should not be discarded either without certainty of the considered issue, as this distorts the true active-inactive distribution that needs to be preserved (note that calculating decoys from actives would in general lead to a different distribution). For this reason, such negative data is valuable and can only be approximated by decoys. Also, to understand how the considered types of methods compare on a new benchmark, diverse and broad methods must be tested. Otherwise, conclusions might quickly found to be incorrect[95,96]. There are now many studies that have investigated best practices in benchmark generation and use[23,29,31,97–99].

### 5 Challenges from the researcher's bias

### 5.1 Researchers with expertise in an AI topic

AI experts tend to think that any problem can be cracked by using the right learning algorithm. Thus, AI for Science initiatives have proliferated to solve scientific problems with primarily computer science approaches. Shortcomings in understanding domain knowledge requirements have typically led to overhyping AI applications, which also represents a challenge to make progress. There is no doubt that AI is having a tremendous impact on drug discovery[15,16,22,100,101] by leveraging not only AI innovations, but also the continuous data increase and maturity of relevant multidisciplinary expertise. Nevertheless, it is often unclear to which extent this is the case and which of the AI models that excel in benchmarks also have prospective value. This is partly due to most AI for Science projects being guided by problem-misaligned benchmarks and sometimes results not being reproducible[102–104].

Here we have discussed realistic data traits that currently pose enormous challenges to learning algorithms. AI experts must collaborate with experts in other disciplines to have a chance of making a real impact on drug discovery. This requires identifying mutually beneficial collaborations rather than service from one expert to another, contributing to drive project direction rather than arriving only in time for post-mortem analyses[105], and learning the basics of the other disciplines to enable fluent communication[106–108]. Such collaborations are critical to learn how to design AI projects that are able to make valuable prospective discoveries.

**5.2 Researchers with expertise in a drug discovery topic**

Conversely, experts in various aspects of drug discovery (e.g. medicinal chemists, computational chemists) tend to be overcritical of AI applications[22,109], which also constitutes a challenge to overcome. This probably boils down to fear of becoming less relevant or even redundant, but this defensive attitude actually makes such outcomes more likely to happen in the future. Traditionally, many drug discovery experts have substituted equal-to-equal collaborations with AI experts by having early-career scientists with an AI background (e.g. PhD students) in their own labs despite not having the minimal expertise to supervise them properly. Among other things, this has led to suboptimal AI applications or even being frustrated by their negative results.

These experts must educate themselves to become at least conversant in AI by taking introductory ML courses (e.g. https://developers.google.com/machine-learning) and reading easy-to-digest primers in ML[110]. They have to revisit their knowledge in the light of more data and its more advanced analysis. After all, reproducibility studies show an alarming level of fatal errors in relevant biological studies[111–114], certainly much larger than in AI studies. In collaborations, they have to share the leadership of projects with AI experts. Importantly, not every problem is currently better served by complex AI models, hence finding the niche of AI and non-AI models across problem instances is also an important outcome of these collaborations in which drug discovery experts play a key role.

Model interpretability is an aspect that drug discovery experts regard as crucial for gaining their trust and accelerating their adoption of AI models[115,116]. Interpretability can be defined as understanding how the property predicted by the model changes with the different characteristics of the data instances (i.e. changes in the values of the features). For example, the property could be the activity of the considered molecule on a given target and the features could correspond to its chemical groups. Here medicinal chemists would typically like to interrogate the model to find out which groups are most important for activity in order to optimise the activity of the molecule by changing those groups using their knowledge.

Unfortunately, models with higher interpretability typically come at the cost of being less predictive. Another issue, often overlooked, is that they can only possibly find out which groups are most important for the activity predicted by the model, not the measured activity. Therefore, interpreting a model only makes sense if it is highly predictive, i.e. if measured and predicted activity are highly correlated, which is usually not the case. Given these challenges, drug discovery experts should consider trusting AI models by means other than fully understanding them. Just in much the same way that most people entrust their lives to cars that they know how to operate, but whose operation they cannot effectively interpret, build, repair or test.

On the other hand, competitions must encourage progress, not triumph, and be inclusive, not just say they are. For instance, CACHE (https://cache-challenge.org/) applicants are evaluated by other applicants who are mostly not AI experts. Unsurprisingly, these rules strongly reduce the participation of AI experts. Furthermore, CACHE's first competition, which aimed at discovering LRRK2 binders, was advertised with statements like "When it comes to AI and drug discovery it can be hard to separate the science from the hype", which it could also be said of non-AI methods such as those based on molecular modelling. Note that LRRK2 had no known binders, which is the worse-case scenario for AI models as there is no target-specific data to train them. This is of course a valid scenario. However, if the hype of AI for drug discovery is so intense, why do you need to implicitly exclude AI experts and select the most challenging target type for AI models? Even if the competition was balanced, any conclusion would be at best restricted to LRRK2. By contrast, the outstanding usefulness of AI models, here a structure-based ML scoring function, has already been proven prospectively across 318 targets[15]. It is worth mentioning that other competitions have also shown the advantages of taking a ML approach to these problems (e.g. Wei's lab entry in the D3R Grand Challenges[117]).

**5.3 Researchers without experience in prospective applications**

Prospective validations are a univocal way to demonstrate the value of a computational method. For instance, there is a plethora of studies reporting the discovery of molecules with previously unknown activity in a range of targets[15,19,124,47,68,118–123]. It is hence surprising that the vast majority of AI experts reporting outstanding retrospective results are not reporting any of such discoveries using their AI models. This criticism is also directed to drug discovery experts who have not led any successful application using their alternative non-AI models. This is particularly shocking when these experts are academics in departments with wet-lab capabilities or researchers in pharmaceutical companies (*in vitro* validations can be easily carried out on public molecules and targets). How can these researchers be so sure of their published benchmarks, criticisms and models if they have not led to substantial discoveries?

Prospective applications of computational methods are a humbling and learning experience. AI experts can learn that endlessly modifying your model until it clocks top performance on a benchmark is very different from having a single shot to submit your predictions to your wet-lab collaborator for in vitro assessment. Drug discovery experts can learn that the assumptions on which retrospective studies are based are more a reflection of their educational biases than there are an accurate embodiment of the factors influencing prospective success. We encourage the publication of negative results, given their high value for this research area. When only studies with positive results are published, inexperienced stakeholders are led to believe that the method is effective in every case. A refreshing exception is a recent work by Atomwise[15], where their structure-based ML scoring function (AtomNet) was prospectively evaluated over 318 targets. Despite including hard targets (no ligand-bound crystal structure, no known ligands) and only in vitro testing molecules chemically dissimilar from known target actives, submicromolar actives were discovered from dose-response experiments in about 60% of the targets[15].

Of these 318 targets, 22 targets were internal efforts of Atomwise and the remaining 296 targets from their academic collaboration programme (Artificial Intelligence Molecular Screen; AIMS). In AIMS, Atomwise partnered with 482 academic labs and screening centers, from 257 different academic institutions across 30 countries. Essentially, Atomwise used AtomNet to identify molecules with high predicted potency against the target and paid Enamine to synthesise them, whereas the academic partner/s tested the synthesised molecules on their assays for the target. To achieve substantial hit rates across most targets, they only had to synthesise and test an average of 85 molecules per target[15]. Once the AIMS programme cost is compared to the €234 million the European Lead Factory has required to deliver over 10,000 actives of undisclosed quality (https://elfendofprojectreport.europeanleadfactory.eu/), it becomes clear that public funding bodies should already be funding AIMS-like partnerships between AI experts with a record of successful prospective studies and experts in therapeutic targets. We think that a prospective success record achieved on diverse targets with different wet-lab collaborators is the most convincing one. More so, if the studies are conducted in resource-constraint environments, where only few molecules and targets could be tested *in vitro*.

## 6 Conclusions

In giving our perspective of the challenges in AI for drug discovery, we have taken a data-centric approach. There are multiple data issues which collectively challenge the generalisation of AI models, the ones we reviewed (bias, inconsistency, skewness, irrelevance, small size, high dimensionality) plus some other we did not have the space to cover (e.g. data sharing).

We have seen that, even if perfect data was possible, AI modelling would not only be limited by methodology shortcomings, but also by unrealistic benchmarks, inadequate performance assessment, lack of balanced expertise and/or expertise-dependent human bias.

## 7 Expert opinion

AI is transforming drug discovery. There is however much confusion about which AI models are most effective for specific problems and how much they outperform non-AI methods. The current reality is that, still in many problem instances, AI provides only marginal improvements or does not even provide the best solution. In addition to overlooking this fact, many editors, funders, investors and researchers in AI for drug discovery consider that all you need to make progress is a new deep learning architecture, more GPUs and larger training sets.

To help bring clarity to this question, we have given an overview of what can go wrong in practice with AI for drug discovery. It is important to be aware of these challenges so that models can be designed to overcome them. Here we have reviewed a range of strategies that have been shown to be able to mitigate the data-centric challenges. We have also presented methods for quantifying predictive uncertainty as a promising route to complement these efforts, although much work is needed to make progress in this front.

Model performance evaluation is also fraught with obstacles, which limits what can currently be achieved in AI for drug discovery. It requires a solid understanding of the relevant concepts and metrics, complemented by benchmarks that can anticipate which models will generalise better to distinct molecules, therapeutic targets and phenotypic endpoints.

Another challenge is missing essential expertise, either AI or domain knowledge, when tackling a problem. Furthermore, we explained how human bias, whether from AI experts or drug discovery experts, constitutes another challenge that can be alleviated with prospective studies. In particular, progress could be boosted if more researchers directed their collaborative efforts to understand how their models could be developed to work better prospectively.

Progress towards overcoming these challenges will further reduce the costs and timelines of bringing new drugs to patients, liberating resources that could be used to increase the number of discovered drugs as well. In addition, multi-disciplinary collaborations using improved AI models will eventually enhance the discovery of novel therapeutic targets and the selection of patients for a given drug to increase therapeutic benefit while reducing harmful side effects.

> **Article Highlights**
> - The impact of distribution shifts can be investigated with suitable training-test data splits.
> - Optimal trade-offs between reducing data inconsistencies and increasing the accuracy and scope of AI models trained on these datasets can be reached.
> - To tackle skewed data (class imbalance in classification), supervised learning following oversampling of the majority class (e.g. by generating decoys), active learning and semi-supervised learning have successfully been used.
> - Feature selection techniques help to analyse irrelevant and/or high-dimensional data.
> - When the dataset is small, methods for self-supervised learning and transfer learning have been found effective.
> - Other recommendations to overcome current limitations are: developing uncertainty quantification methods, building realistic benchmarks using task-aligned performance metrics, reflecting on our own biases and prioritising the prospective application of AI models for drug discovery.

## Author contributions

P.J.B. conceived this article and wrote it with the assistance of G.G., S. H-H. and C.P.

## Funding

The authors are supported by the Chulabhorn Royal Academy of Thailand, the Marie Skłodowska-Curie Actions under Horizon 2020 (grant no. LCII_PA4887), the National Council of Sciences and Technology of Mexico, the Royal Society (award no. RSWF\R1\221005) and the Wolfson Foundation (award no. RSWF\R1\221005).

**Acknowledgements**

We thank Qianrong Guo and Zerui Li for their feedback on an early draft of this paper.